# High-toughness/low-friction ductile epoxy coatings reinforced with carbon nanostructures

Anderson O. Okonkwo[1], Pravin Jagadale[2], John E. García Herrera[3], Viktor G. Hadjiev[4], Juan Muñoz Saldaña[3], Alberto Tagliaferro[2], Francisco C. Robles Hernandez[1*]

[1] Mechanical Engineering Technology Department, College of Technology, University of Houston Texas 77204-4020 USA.

[2] Department of Applied Science and Technology Politecnico di Torino, C.so Duca degli Abruzzi, 24-10129, Torino-Italy.

[3] Centro de Investigaciones y Estudios Avanzados del Instituto Politécnico Nacional, Unidad Querétaro, Libramiento Norponiente #2000, Real de Juriquilla. C.P. 76230, Mexico.

[4] Texas Center for Superconductivity and Department of Mechanical Engineering, University of Houston, Houston, TX 77204, USA.

* Author to whom correspondence should be addressed to: e-mail: fcrobles@uh.edu, tel. 1-713-743-8231

**Abstract**

We present the results of an effective reinforcement of epoxy resin matrix with fullerene carbon soot. The optimal carbon soot addition of 1 wt. % results in a toughness improvement of almost 20 times. The optimized soot-epoxy composites also show an increase in tensile elongation of more than 13 %, thus indicating a change of the failure mechanism in tension from brittle to ductile. Additionally, the coefficient of friction is reduced from its 0.91 value in plain epoxy resin to 0.15 in the optimized composite. In the optimized composite, the lateral forces during nanoscratching decrease as much as 80 % with enhancement of the elastic modulus and hardness by 43 % and 94%, respectively. The optimized epoxy resin fullerene soot composite can be a strong candidate for coating applications where toughness, low friction, ductility and light weight are important.

# 1. Introduction

Carbon nanostructures such as fullerene [1], nanotubes [2] and graphene [3] have been widely used to reinforce different inorganic matrices [4-9] and polymers [10], thus producing composites with improved mechanical or multifunctional properties. The most common polymer matrices include epoxy resin (e.g. "epoxy") [11], polyester [12], polyvinyl [13] and polyethylene [14]. The use of carbon nanotubes as fillers in polymer composites have led to the improvement of mechanical properties such as strength, toughness, elongation, Young's modulusand wear [13, 15-17]. A positive impact was also obtained on both AC and DC electrical conductivity [15]. Proper functionalization of carbon nanostructures provides further enhancement of the mechanical properties of composites [18]. Chemical interactions between the reinforcement and polymeric matrix may result in further enhancement of the mechanical properties [19]. Carbon particles (e.g. nanotubes or graphenes) can be the key to producing polymeric matrices with multi-functional character for manufacture of lightweight components for advanced applications (aerospace, electronics, automotive etc.). However, the literature highlights that a serious limitation in this type of composites is represented by the inefficient dispersion of the nanotubes in the host matrix [20, 21].

In the present work, a methodology to reinforce an epoxy resin with nanostructured carbon soot filler using an ULTRA-TURRAX digital high-speed homogenizer system is exploited. The mechanical improvements reported here are well beyond those observed so far with other reinforcing nanofillers such as nanotubes. The improvements include a sizable increase in toughness, hardness and ultra-plastic behavior as well as lowering of the friction coefficient of composites. Moreover, all improvements occur all together in a high quality composite with established optimized carbon soot loading of 1wt. %. The simplicity of technological operations and the low cost filler used to achieve such results provide a further advantage. It is expected that this technology can be implemented, with minor modification, for mass production of materials for coatings where toughness, plasticity, hardness and reduced friction are important.

## 2. Material and Methods

The fullerene carbon soot is produced by the Kratschmer method [22] and is the byproduct obtained after the purification of fullerene. The soot used in the present work has less than 1 wt. % fullerenes ($C_{60}$ and $C_{70}$).

The thermoset epoxy polymer used in the analysis is made with two parts: epoxy resin and a cross linker. Epoxy resin (Epilox® T 19-36/700) is colorless, low viscosity (650-750 mPa.s at 25 °C) with (density of 1.14 g/cm$^3$). Its main components are Bisphenol A and Glycidyl ether. The cross linker (Epilox® H 10-31) is a colorless liquid, low viscosity (400-600 mPa.s at 25 °C), with density of 1 g/cm$^3$. It is formulated with 3-aminomethyl-3, 5, 5-trimethyl-cyclohexylamine and Benzyl alcohol.

Resin (T 19-36/700), cross linker (H 10-31) and soot as filler were thoroughly mixed in specific ratios with mechanical stirring (20,000 RPM for 2 minutes). A subsequent sonication step (ultrasonic frequency 37 KHz for 15 min) followed by degassing in vacuum was performed to make sure that all trapped bubbles were completely removed. Before the polymer cures it was poured into the mold. Epoxy polymer and composite were thermally cured at 70 °C for 4 hours in an oven. Samples were prepared with two different CS concentrations: 1 and 3 wt. %.

X-ray diffraction (XRD) was carried out using a D5000 SIEMENS diffractometer, with a Cu tube and a characteristic $K_\alpha$ = 0.15406 nm operated a 40kV and 30 A. The scanning electron microscopy (SEM) observations were carried out using two field emission SEM's. One is a FEI XL-30FEG and the other is a FE-SEM Zeiss Supra 40 connected to an Energy dispersive X-ray spectroscometer (EDS-Oxford Inca Energy 450). The high resolution transmission electron microscope observations (HRTEM) were carried out on a Jeol 2000FX operated at 200 kV. The HRTEM images were analyzed using Digital Micrograph 3.7.1 software. X-ray photoelectron spectroscopy (XPS) was conducted on a Physical Electronics XPS Instrument Model 5700, operated via monochromatic Al-$K_\alpha$ X-ray source (1486.6 eV) at 350 W. The data analysis was conducted with Multipak™ software, and the Shirley background subtraction routine had been applied throughout.

The raw powder was analyzed before and after calorimetric analysis with Raman using a Renishaw Micro Raman system with green laser line (wavelength: 514 nm) equipped with a CCD detector. The microscope used a 50X objective lens to focus the laser beam on the sample surface, and the size of the focused laser spot had a diameter of a few micrometers. The composites were analyzed in a confocal micro-Raman XploRA™, Horiba JY using a Raman excitation green laser of a 532 nm.

For the characterization of mechanical properties, a defect-free region of the sample surface was selected by atomic force microscopy imaging prior to the indentation test. Indentation measurements were conducted using an Ubi1 instrument (Hysitron, Minneapolis). The machine compliance and the area function of the tip were calibrated before the indentation test using a fused silica sample (ASMEC, Germany). The loading and unloading segments in trapezoidal three-segment load function were each completed over a time of 30 s, irrespective of the maximal load ($F_{max}$). $F_{max}$ was kept constant for 30 s. A reference sample of polycarbonate (ASMEC, Germany) was additionally measured to test the calibration of the device.

A set of 36 indents was carried out in a symmetric matrix spaced with a maximum load of 180 µN, where each indentation imprint was separated by at least 4 µm to avoid the influence of the stress fields around the indents; load-penetration curves were recorded in each measurement . A 60 s delay at zero loads was established before and after each indent for thermal drift determination. The hardness is defined as $H_{IT} = F/A_c(h_c)$, where $F$ is the applied load and $A_c$ is the contact area, which is itself a function of the contact depth ($h_c$), as calculated by the Oliver and Pharr Method [23]. For the reduced elastic modulus, the following equation was used:

$$\frac{1}{E_r} = \frac{1-v_i^2}{E_i} + \frac{1-v_s^2}{E_s} = \frac{2}{\sqrt{\pi}} \cdot \frac{\sqrt{A_c(h_c)}}{S} \tag{1}$$

where $E$ and $v$ are the Young´s modulus and Poisson´s ratio and the subscripts $i$ and $s$ refers to the indenter and the sample, respectively. The contact stiffness, $S = dF/dh$ is estimated from the first part of the unloading segment of the load-penetration curve. It is worth mentioning that the viscoelastic effects on the determination of reduced elastic modulus were neglected in this work but deserve to be determined in a separate contribution.

At least 5 nanoscratch tests were performed on each sample using a Knoop tip in an IBIS-UMIS nanoindentation device in a steady load mode. The load was varied between 5 to 9 mN with 1 mN increments. Each scratch test was performed over a length of 500 µm, recording continuously the lateral force as well as the friction coefficient through a force sensor LVDT. A pre-scan was made for slope correction, which is done with the closed loop PZT direct acting normal force sensor that keeps the load for curved or sloping surfaces.

## 3. Results and Discussion

A summary of carbon soot characterization results is presented in Figure 1. The SEM micrograph in Figure 1a reveals a fluffy morphology of carbon soot. The particles are nanostructured with amorphous and/or short distance ordered that is evident in Figure 1b (HRTEM). The XRD results presented in Figure 1c are dominated by the (002) reflection of graphitic carbon and the x-ray signature of $C_{60}$ fullerite (molecular crystal) particles. The Raman spectra in Figure 1d corroborates the XRD findings of graphitic structures with short lateral dimensions.

The TGA analysis demonstrates that CS is stable to temperatures of approximately 350 °C with a weight loss lower than 3 wt. %. Another 4 wt.% is lost at temperatures below 70 °C and it is attributed to organic residue and moisture since the soot was tested in as purchased condition. The weight loss of the carbon soot during heating to 700 °C is an additional 83 wt. %. We attribute the above weight reduction to the oxidation of the amorphous material first, followed by oxidation of the short-order graphitic structures. The remaining 10 wt. % was characterized by Raman showing comparable spectra to that observed in the raw material. The Raman results indicate that the remaining carbon is graphitic and nanostructured.

The carbon soot was analyzed by EDS and XPS and the results are presented in Table 1. Both methods found soot to be comprised only of carbon and oxygen. The predominant grain size was calculated using the following relationship [4]

$$L_a(nm) = 2.4 \times 10^{-10} L_{las}^4 (I_G/I_D) \quad , \tag{2}$$

where $L_{las}$ = 638 nm is the excitation laser wavelength, and $I_G$ and $I_D$ are the intensity of the Raman *D* and *G* bands, respectively. The XPS analysis shows a majority of carbon and the balance is oxygen.  This observation was also confirmed by EDS measurements. The results of CS characterization are presented in Table 1.

The Raman D and G bands shown in Figure 1d are typical for $sp^2$ rich carbon materials [24]. The G band is due to the symmetric $E_{2g}$ carbon vibrational mode, allowed by Raman selection rules, whereas the D band is a product of defect-induced Raman scattering involving carbon vacancies, functional carbon-oxygen groups, and boundaries of nano-sized graphite particles. The second-order Raman 2D and D+G bands involving two phonons appear only in $sp^2$ material with translational order [24].  The BET results indicate that the surface area of the soot is 161 $m^2$/g and its density 1 $g/cm^3$.  We conclude that the soot is in the form of spheres composed of a mix of amorphous and graphitic structures with short range order and a limited amount of non-graphitic material.

Figure 2 shows the SEM images of the epoxy and the composites with 1 and 3 wt. % soot.  The surface morphologies of epoxy and composites are markedly different; the latter reveals clearly the embedded spherical nanostructures (100-150 nm) in the polymeric matrix. Figure2d shows that the epoxy Raman fingerprints are seen in the three investigated samples. The characteristic graphitic carbon band are clearly discernible and show little deviation from those observed in the raw soot (Figure 1d), indicating no apparent damage or modification of the soot. Therefore, from those results, we conclude that no massive chemical interaction between the epoxy and the soot takes place; instead, the interactions are through van der Waals forces. As expected, the intensity of the carbon response increases with the amount of soot.

The tensile testing results are presented in Figure 3. The epoxy sample shows a stress-strain curve characteristic of a brittle material with almost no plastic deformation, and an ultimate tensile strength (UTS) of 17.7 MPa. Its measured Young's modulus is 1.8 GPa.  The composite with 1 wt. % of soot presents a slight increase of 5 % in UTS (18.6 MPa) whereas with 3 wt. % soot the strength is comparable to that of the epoxy.  The Young's modulus for both composites is approximately 2.0 GPa, resulting in a 13.3 % increase with respect to pure epoxy.  The yield strengths are 13.2 and 12.3 MPa for 1 and 3 wt. % soot additions, respectively.  Due to the brittleness in the epoxy, there is no identifiable yield point for it, while the strength at failure is 17.5 MPa.  The most important result, however, is the large increase of plasticity of the composites, particularly in the 1 wt. % composite, reaching 13.2 % elongation at a stress of 14.1 MPa.  The composite with 3 wt. % soot addition had a maximum elongation of 7.0 % at a stress of 14.3 MPa.

The absorbed energy during tensile testing or toughness presents significant improvements in the composites as compared to the results of the epoxy.  The pure epoxy does not have modulus of resilience because it fails under brittle regime; hence, it lacks of yield point.  On the contrary, the composites have a resilience of 9.1 and 7.1 MPa for the composites with 1 and 3 wt. % of carbon soot, respectively.  The toughness of the pure resin is 9.6 MPa and the corresponding values for the composites are 186.2 and 107.2 MPa for the 1 and 3 wt. % soot.  Therefore, the resilience strongly increases by inducing ductility in the epoxy. The corresponding toughness improvements are 1845 % and 1020 %.  The improved mechanical properties make the composite an appealing material for coatings that undergo significant plastic deformation.  This composite is a light weight and may be an ideal for coating to absorb impacts (e.g. blast).

These results also suggest that the elongation of the epoxy composites can be tuned by varying the carbon soot loading.  We have found, however, that the loading of 3 wt. % is the limit of homogeneous dispersion of soot in the epoxy, and further increase of the filler content deteriorates the dispersion. Although various filler loadings were tested, we present in more details the composite with 1 wt. % soot because of its best performance. The 3wt. % loading results are given to highlight the effect of the onset of dispersion problems. It is worth noting that both composites show upper and lower yield strengths similar to those observed in low carbon steels [25]. Most probably this behavior is due to the cross linking inhibition due to the presence of 3D structured soot particles. In other words, the composites elastic and plastic behaviors are markedly different from that of the epoxy as a result of a reduction in the cross linking density in the epoxy matrix.

Figure 4 shows SEM micrographs for the composites. The spherical soot nanoparticles are discernible in both composites. The soot particles are limiting the cross linking of the epoxy resin that results in the enhancement of plastic region when the composite is under tensile stress. This mechanism is more effective in the composite with 1 wt. % soot. Higher density of particles may act as stress concentrator restricting the plastic enhancement and increasing the stress in neighboring regions less rich in filler. This is evident in the composite with 3 wt. % soot. Figure 4c shows how soot particles allow cracking control under plastic conditions acting as a crack stopper. Thus, the voids contribute to the increased plasticity. The particle dispersion in the 1wt. % soot composite is more effective because of the lack of agglomeration. Nevertheless, in both composites the plastic behavior is clearly present. In contrast, in the pure epoxy cracking initiates and propagates in the absence of plastic deformation.

The nanoscratch testing shows that addition of different amounts of soot into the epoxy have remarkable differences on its tribological behavior. A summary of the nanoscratch testing results is presented in Figure 5, where the scratches along the surface of the investigated samples are evident. Each scratch was obtained using a constant load. Loads were varied in the range 5 to 9 µN. Figure 5a-c displays the nanoscratch test results for the epoxy and composites reinforced with 1 and 3 wt. % soot, respectively. The load increases from the first to the last scratch, as indicated in the figure. The deeper and more defined scratches are in the epoxy followed by the 3 wt. % soot composite. The 1 wt. % soot composite shows the least damage.

Figure 5d depicts the lateral forces during the nanoscratch test. A steady state is reached in all tests such that the forces are essentially constant. The 1 wt. % soot composite has the least resistance to the nanoscratch. This is a consequence of a lubricity effect ongoing on this composite. The composite with 3 wt. % soot shows an increase in the lateral forces. One possible explanation is that agglomeration of the soot particles could contribute to a deteriorating stress concentration effect in the composite. The pure epoxy sample exhibits the highest lateral forces (Figure 5b-c). Besides the increased plastic behavior in tension, the tribological properties of the composites present distinct advantages over the pure epoxy with a decrease in friction forces and increased lubricity.

The coefficient of friction (CF) as a function of applied normal load is presented in Figure 6. The CF in the epoxy is reduced from 0.91 to 0.59 when the load is increased from 5 to 9 $\mu N$. For the 1wt. % soot composite, CF varies little with the test load, being consistently in the range of 0.15 to 0.16. This is equivalent to 83% reduction in the CF when compared to the epoxy under 5 $\mu N$ load, and 73% under 9 $\mu N$ load. On the other hand, for the 3 wt. % soot composite, the CF decreases with load from 0.56 (5 $\mu N$) to 0.39 (9 $\mu N$).

The elastic modulus and nanohardness test results are presented in Figure 7. A direct comparison shows that both properties are improved in the composites relative to the epoxy. The data scattering (standard deviation) in both composites is also reduced. However, as the nanomechanical measurement tests small subsurface volumes of the sample, the less homogeneous dispersion of carbon soot in the 3 wt. % case leads to larger data scattering. This is an indication of the higher homogeneity in the material and effective reinforcement by the carbon soot.  The composites show improvements in the average elastic modulus of 9.9 and 16.7 %, while the corresponding improvements in hardness are 16.6 and 28.6 % for the composites containing 3 and 1 wt. % soot, respectively.

In summary, the addition of fullerene soot leads to a relevant increase of resilience, toughness, ductility, lubricity and hardness while decreasing friction, hence making the epoxy matrix ductile.  This particular type of soot is resistant to temperatures of up to 329 °C in air, which make it also suitable for use in fire retardant applications.  The presence and morphology of the fullerene soot in the composite is clearly identifiable by means of Raman spectroscopy and microscopy. During tensile testing, the Young's modulus of the epoxy is preserved in the composite; therefore, we presume that the epoxy (matrix) does not suffer major molecular changes other than the local inhibition of cross linking along the soot particles.  This is confirmed by the similarity in the Raman spectra of epoxy and the composites.  The remarkable improvements in toughness, resilience, hardness and lubricity make this composite ideal for coating applications.

We relate the increase of elongation of the epoxy composites to the reduction of local cross linking by the soot particles. The combination of size of the spherical soot particles (<160 nm) and loading of 1 wt. % results in an optimal homogeneous dispersion of large surface area particles within the epoxy (Figure 4).  During the tensile tests, the stress is mainly carried by the epoxy matrix until reaching the elastic limit. Afterwards, the filler starts to play its key role by limiting crack propagation. The dispersion of the soot particles in the 3 wt. % composite is not as effective as in the 1 wt. % case, leading to soot agglomerations acting as stress concentrators. Nonetheless, even with up to 3 wt% soot, the plastic behavior during cracking is present in the composites. There is, however, a specific balance between particle size and particle concentration that makes the effect the most significant. The 3 wt. % loading was the limit of dispersion of soot in the epoxy, larger amounts result in excessive soot agglomeration limiting the ductility of the composite.

During the scratch test, the lateral forces in the epoxy compared to those in the composites show marked differences. Larger lateral forces are observed in the epoxy while the lowest are found in the composite with 1 wt% soot, having a force reduction of approximately 80 %. We attribute that to a potential lubrication mechanism occurring in the composites while dragging the soot particles with the tip indenter. In addition, the composite with 1 wt% carbon has the highest hardness and reduced elastic modulus with the narrowest scatter of the data.

## 4. Conclusions

The addition of fullerene soot in epoxy results in an overall improvement in resilience, toughness, strength, hardness, coefficient of friction and modulus of the resulting composites. The most significant result is the change in failure mechanism from brittle to ductile during tensile testing, along with a sizable development of resilience and clear improvements in toughness of approximately 20 times. The ductile behavior is attributed to the soot particles behaving as local inhibitors of the cross linking and crack stoppers in the epoxy, allowing controlled cracking under a plastic regime. A remarkable increase of elongation is observed from 0.7 % in the epoxy to more than 13 % in the composite with 1 wt. % soot. In the same composite, the coefficient of friction is reduced by 83 %, becoming almost independent of the applied load. Elastic modulus and hardness are enhanced by almost 50 % and 94 %, respectively. These outstanding mechanical properties of the 1 wt. % fullerene soot epoxy composite make it a very strong candidate for coating applications; for instance, impact or blast absorption.

## Acknowledgements

The authors would like to thank Dr. Martin Wagner from LEUNA-Harze GmbH for providing Epoxy resin as well as Dr. Salvatore Guastella for the FESEM of the composites after tensile testing.  FCRH would like to express his gratitude to the University of Houston and the government of Texas for the Start Up and HEAFS funding. VGH work was supported by the State of Texas through the Texas Center for Superconductivity at the University of Houston.

**Table 1. The carbon soot composition obtained by means of XPS, EDS and Raman.**

| Composition (at %) | | | Grain size (nm) |
|---|---|---|---|
| sp$^2$ C | sp$^3$ C | O | |
| 90 | 4 | 6 | 40 ± 20 |

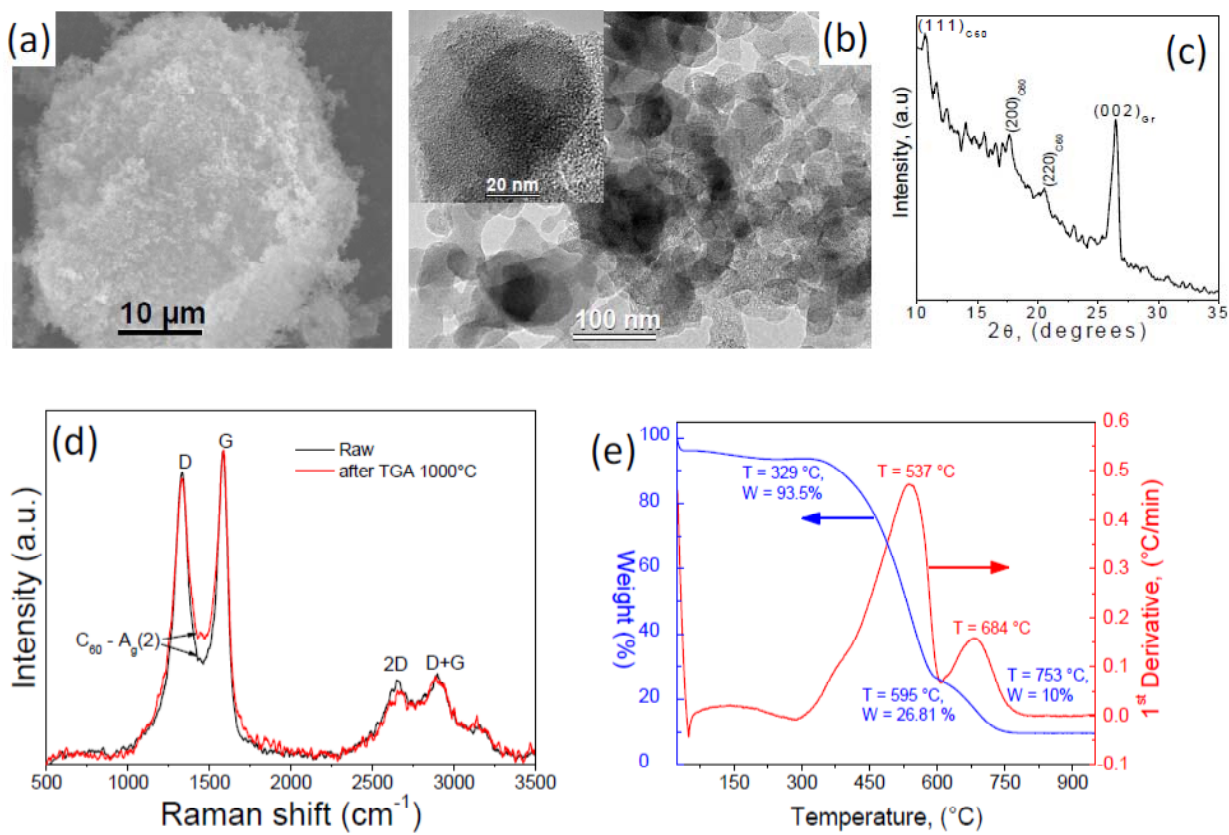

Figure 1. Characterization of soot by means of: (a) SEM, (b) HRTEM, (c) XRD, (d) Raman and (e) TGA. In (d) the $C_{60} - A_g(2)$ refers to a Raman band of fullerene ($C_{60}$). The inset in (b) is a magnified region of one of the soot particles.

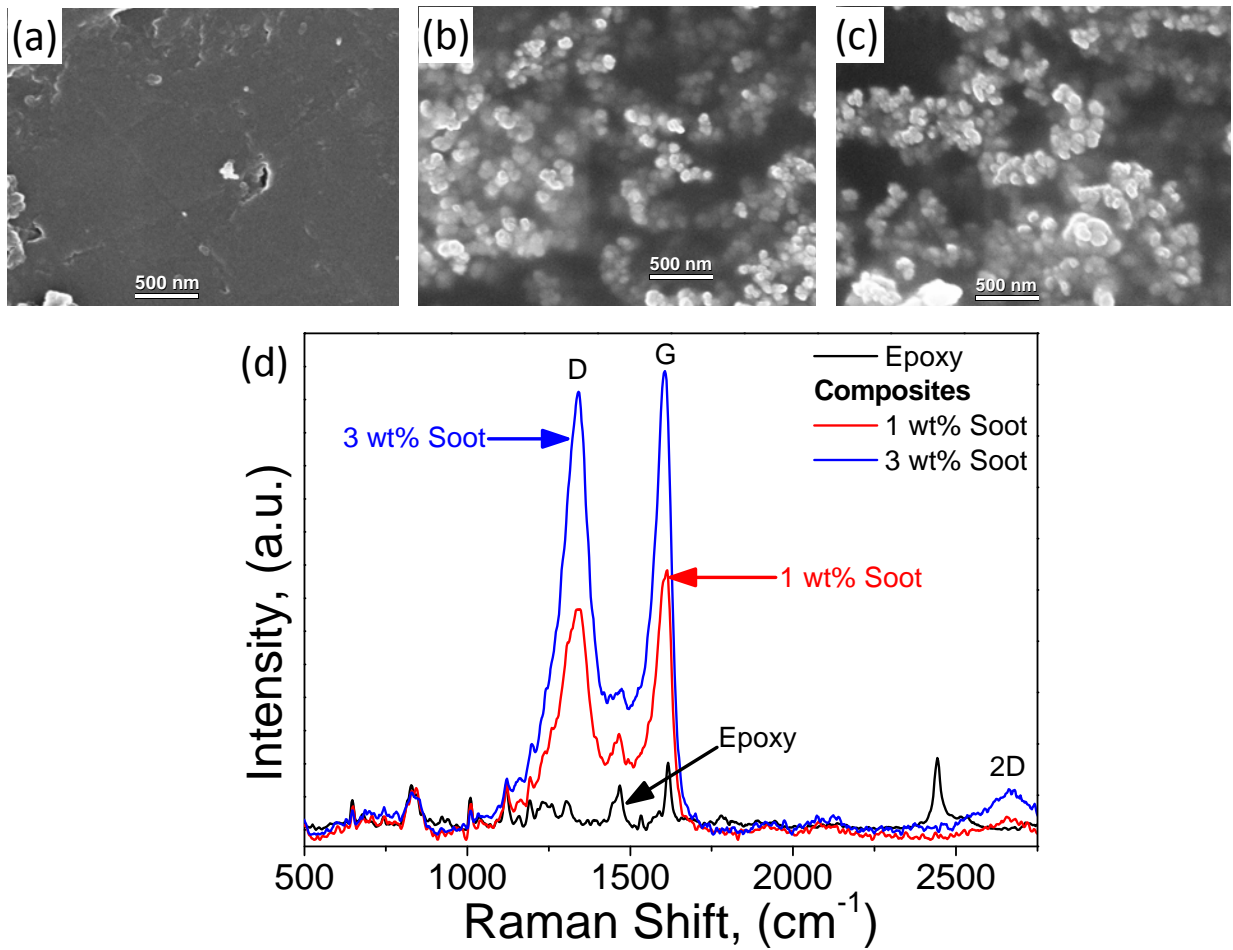

**Figure 2.** SEM micrographs of (a) epoxy and composites with (b) 1 wt% soot, (c) 3 wt% soot and (d) Raman results of the epoxy and composites.

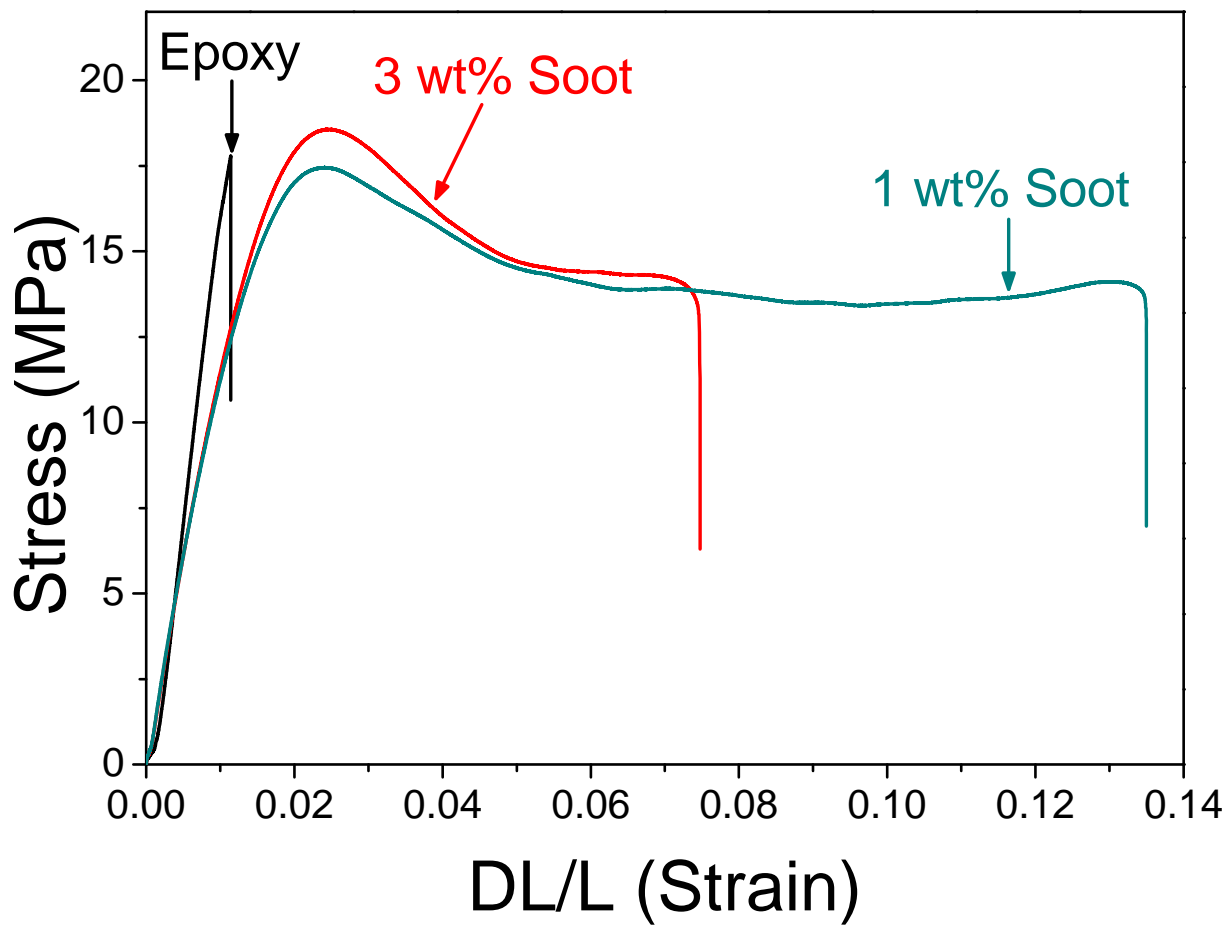

**Figure 3. Tensile testing results of the epoxy resin, and the composites with 1 wt% and 3 wt% carbon soot.**

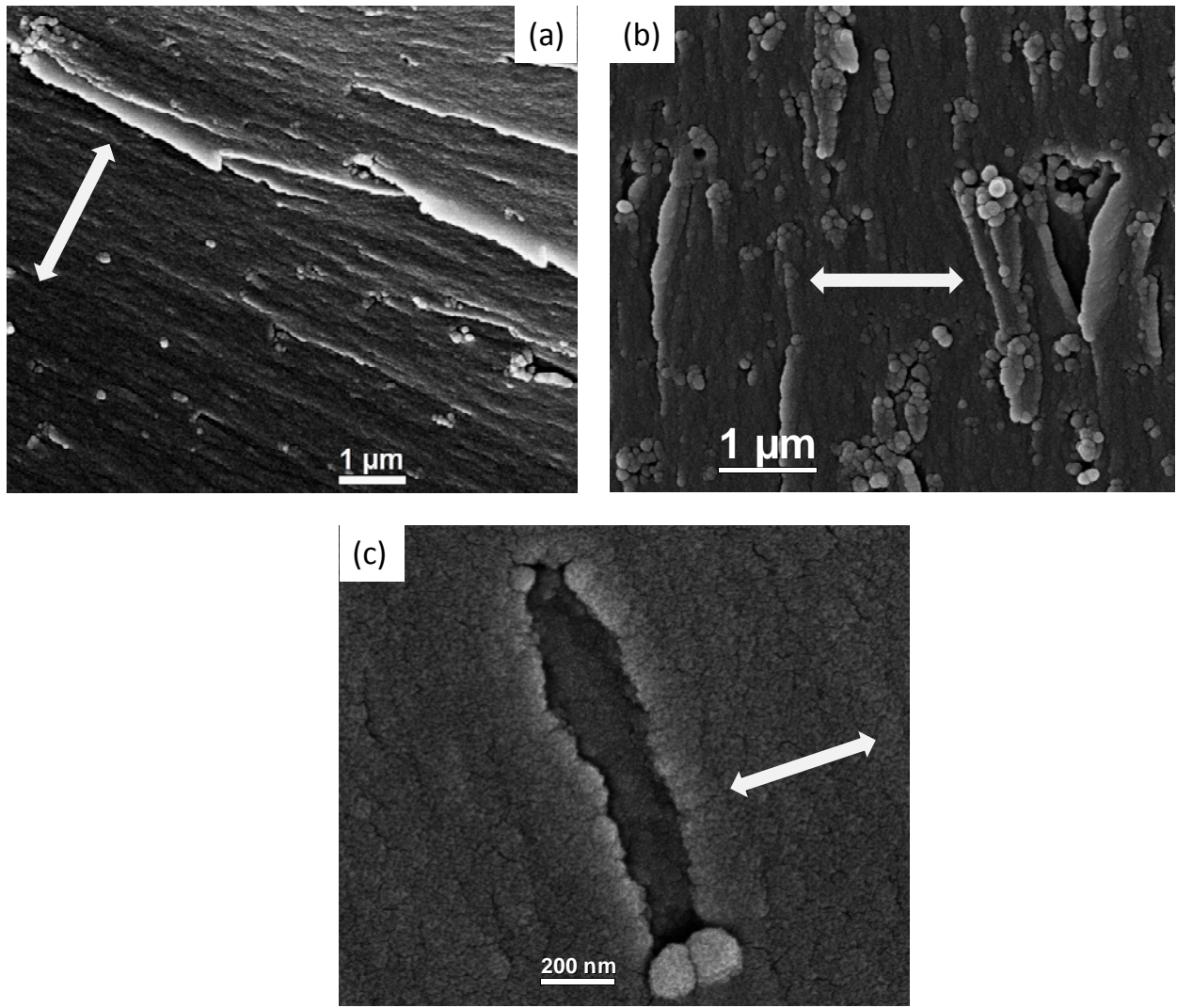

**Figure 4.** SEM micrographs showing the effects of carbon soot particles on the epoxy-soot composite containing (a, c) 1wt% soot and (b) 3wt% soot. The arrow indicates the direction of the applied stress along.

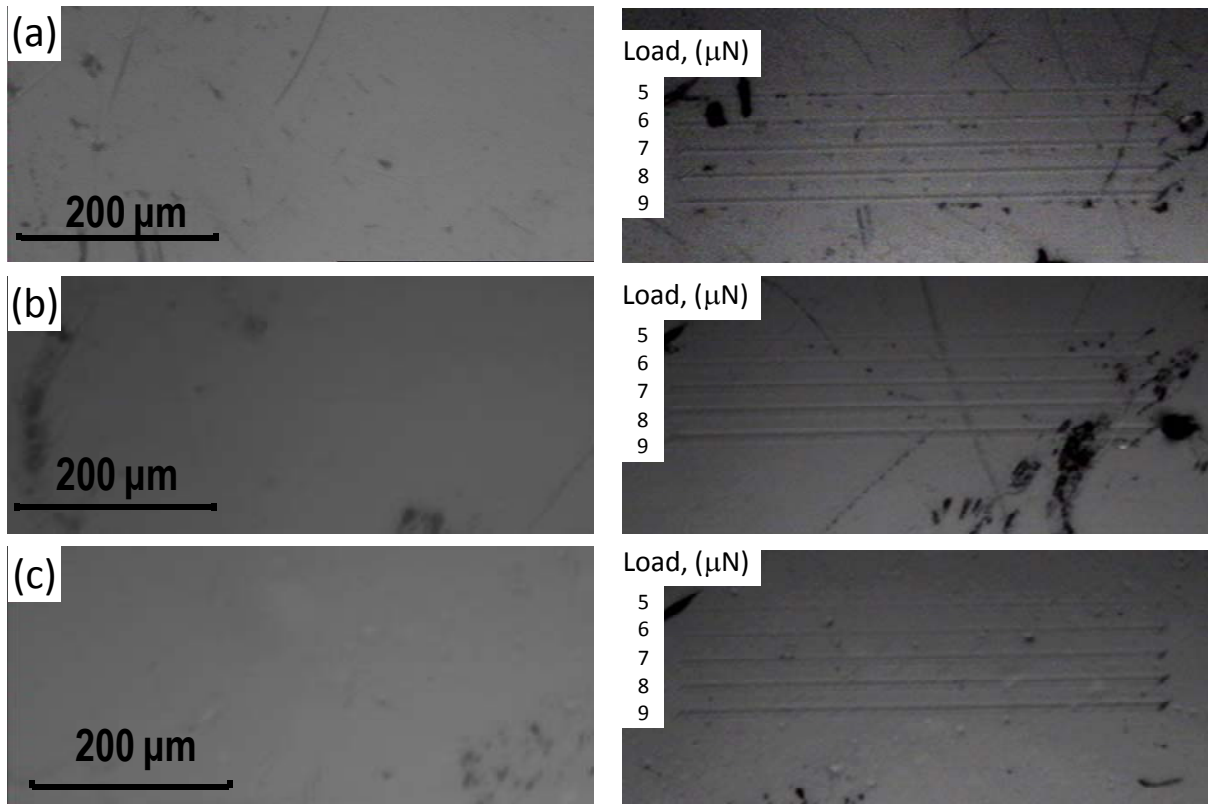

Before                  After

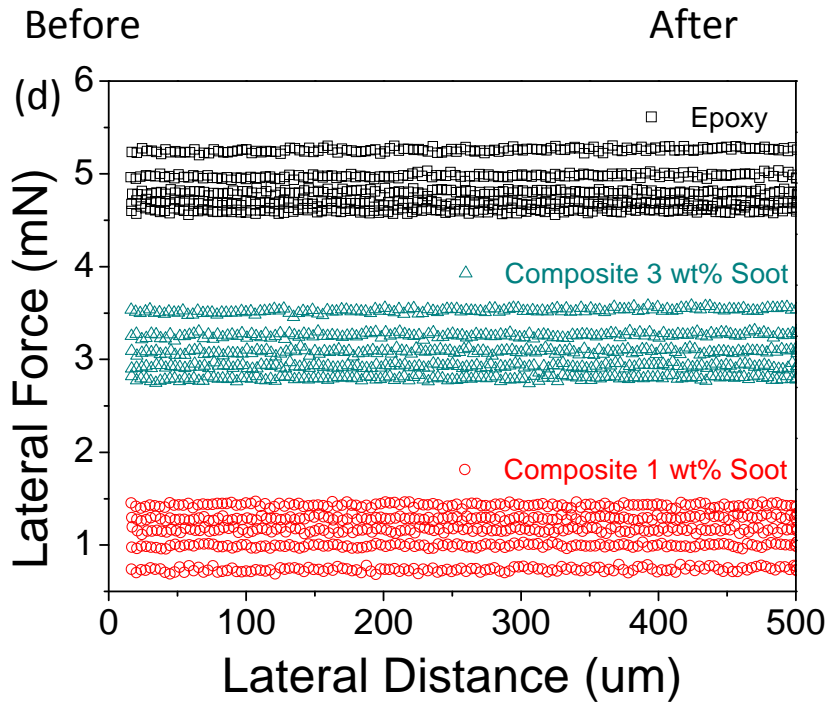

**Figure 5.** Micrographs of the parallel nanoscratch test results on (a) epoxy, and composites with (b) 1 wt%C, (c) 3 wt%C and (d) lateral forces during nanoscratch test using a Knoop tip with loads between 5 to 9 mN with 1 mN increments at 30 μm intervals.

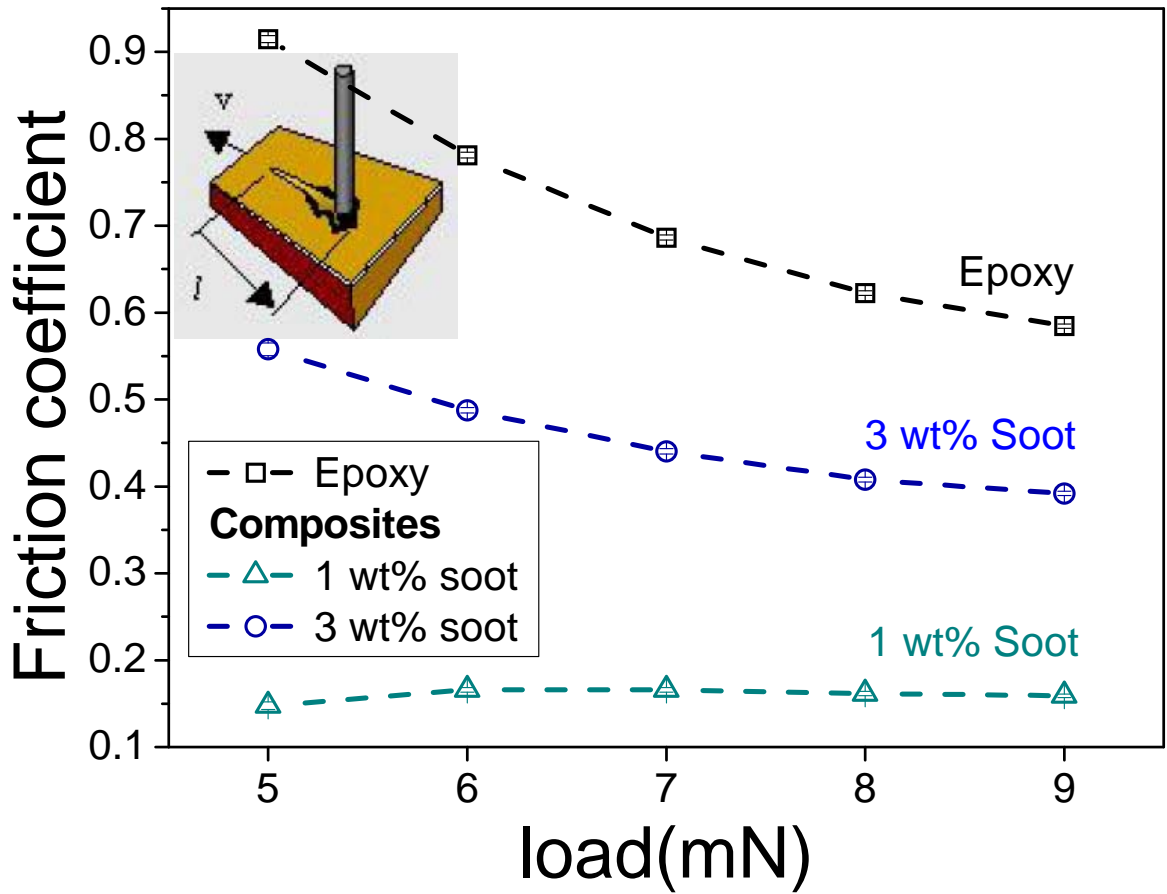

**Figure 6.** Variations of friction coefficient as a function of applied normal load for the epoxy and the composites with 1 and 3 wt % soot.

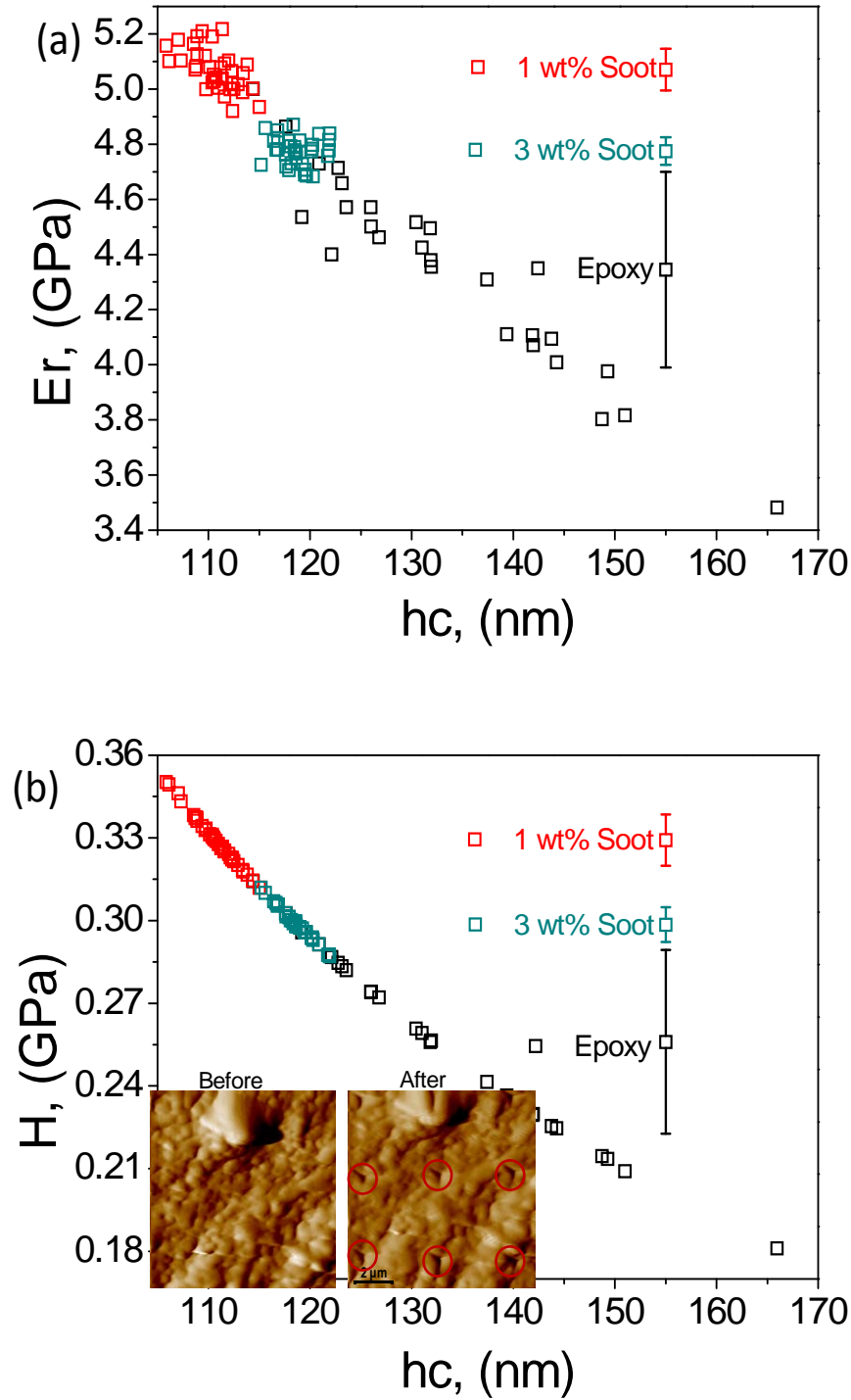

**Figure 7.** Results of (a) modulus and (b) nanohardness for the epoxy and composites with 1 and 3 wt% soot. The scales at about 155 nm indicate the average values and the standard deviation for hardness and modulus respectively. The inset shows micrographs revealing the sample's surface before and after indentation, the red circles identify indentations.